\RequirePackage{booktabs}
\let\clineorig\cline
\documentclass[sn-mathphys,Numbered]{sn-jnl}

\usepackage{multirow}
\usepackage{booktabs}
\usepackage{graphicx}
\usepackage{amsmath,amssymb,amsfonts}
\usepackage{amsthm}
\usepackage{mathrsfs}
\usepackage[title]{appendix}
\usepackage{xcolor}
\usepackage{textcomp}
\usepackage{manyfoot}
\usepackage{algorithm}
\usepackage{algorithmicx}
\usepackage{algpseudocode}
\usepackage{listings}

\usepackage{soul,framed}
\usepackage{xspace}
\usepackage{caption}
\usepackage{titlesec}
\usepackage{makecell}
\usepackage{hyperref}

\makeatletter
\newtheoremstyle{indented}
  {3pt}
  {3pt}
  {\addtolength{\@totalleftmargin}{3.5em}
   \addtolength{\linewidth}{-3.5em}
   \parshape 1 3.5em \linewidth}
  {}
  {\bfseries}
  {.}
  {.5em}
  {}
\makeatother
\theoremstyle{definition}

\theoremstyle{indented}

\titleformat{\paragraph}
{\normalfont\normalsize\bfseries}{\theparagraph}{1em}{}
\titlespacing*{\paragraph}
{0pt}{3.25ex plus 1ex minus .2ex}{1.5ex plus .2ex}

\makeatletter
\newcommand\footnoteref[1]{\protected@xdef\@thefnmark{\ref{#1}}\@footnotemark}
\makeatother

\usepackage{fancyhdr}
\begin{document}

\thispagestyle{fancy}

\title[Sleep Stage Classification using SPD Matrices]{Automatic Classification of Sleep Stages from EEG Signals Using Riemannian Metrics and Transformer Networks}

\author*[1]{\fnm{Mathieu} \sur{Seraphim}}\email{mathieu.seraphim@unicaen.fr}

\author[1]{\fnm{Alexis} \sur{Lechervy}}

\author[3]{\fnm{Florian} \sur{Yger}}

\author[1]{\fnm{Luc} \sur{Brun}}

\author[2]{\fnm{Olivier} \sur{Etard}}

\affil*[1]{Normandie Univ, UNICAEN, ENSICAEN, CNRS, GREYC, 14000 Caen, France}

\affil[2]{Normandie Université, UNICAEN, INSERM, COMETE, CYCERON, CHU Caen, 14000, Caen, France}

\affil[3]{LAMSADE, CNRS, PSL University Paris-Dauphine, 75016 Paris, France}

\abstract{\textbf{Purpose:} In sleep medicine, assessing the evolution of a subject's sleep often involves the costly manual scoring of electroencephalographic (EEG) signals. In recent years, a number of Deep Learning approaches have been proposed to automate this process, mainly by extracting features from said signals. However, despite some promising developments in related problems, such as Brain-Computer Interfaces, analyses of the covariances between brain regions remain underutilized in sleep stage scoring.

\textbf{Methods:} Expanding upon our previous work, we investigate the capabilities of SPDTransNet, a Transformer-derived network designed to classify sleep stages from EEG data through timeseries of covariance matrices. Furthermore, we present a novel way of integrating learned signal-wise features into said matrices without sacrificing their Symmetric Definite Positive (SPD) nature.

\textbf{Results:} Through comparison with other State-of-the-Art models within a methodology optimized for class-wise performance, we achieve a level of performance at or beyond various State-of-the-Art models, both in single-dataset and - particularly - multi-dataset experiments.

\textbf{Conclusion:} In this article, we prove the capabilities of our SPDTransNet model, particularly its adaptability to multi-dataset tasks, within the context of EEG sleep stage scoring - though it could easily be adapted to any classification task involving timeseries of covariance matrices.}


\maketitle

\bmhead{Acknowledgments}

This work was granted access to the HPC resources of IDRIS (Institut du Développement et des Ressources en Informatique Scientifique) under the allocation 2022-AD010613618 made by GENCI (Grand Équipement National de Calcul Intensif), and to the computing resources of CRIANN (Centre Régional Informatique
et d'Applications Numériques de Normandie, Normandy, France).

\section{Introduction}
\label{sec:intro}

Covariance matrices have long been used to analyze groups of concurrent signals, with applications ranging from financial analysis~\cite{finance} to hand gesture recognition~\cite{hand_gestures}.
By construction, they function as descriptors of the relationships between said signals, but also exhibit exploitable mathematical properties, as they usually exist on the non-Euclidean manifold of Symmetric Positive Definite (SPD) matrices~\cite{SPD_book}.
For instance, SPD matrices are commonly used in  structural imagery through Diffusion Tensor Imaging (DTI)~\cite{DTI}, and in brain activity estimation tasks.

Said estimations are often done through the lens of functional connectivity, i.e. the correlations of activity between brain regions~\cite{functional_connectivity}. By construction, correlation and covariance matrices are well suited to such tasks, and have been derived from functional MRI (fMRI) imagery~\cite{fMRI} and electroencephalographic (EEG) signals\footnote{Electrical signals acquired from electrodes located around the brain, often non-invasively.}.
EEG-derived connectivity estimations, in particular, are widely used in the field of Brain-Computer Interfaces (BCI), where they are analyzed through geometry-preserving tools~\cite{BCI1}.

EEG signals are also utilized in sleep medicine as part of the polysomnography (PSG) exam, where they are used to determine the evolution of a subject's sleep over time in order to diagnose sleep disorders.
This has traditionally been a time-consuming process, as it has to be done manually by experts.
in recent years, a number of approaches have been proposed to automate said process (Section~\ref{ssec:SOA_scoring}). However, even though functional connectivity has been shown to be indicative of a subject's sleep stage~\cite{bouchard2019}, most of these focus on analyzing individual signal properties, rather than the interactions between signals. Furthermore, SPD analysis seems largely absent from discussions regarding EEG sleep staging, despite the similarities between this problem and BCI.

In response to this under-utilization of functional connectivity for sleep analysis, we previously developed an approach to the automatic classification of sleep stages from EEG data, predominantly through the analysis of EEG-derived covariance matrices~\cite{CAIP}.
This culminated in \textbf{SPDTransNet}~\cite{arxiv}, a Transformer-based Deep Learning model adapted to analyze sequences of SPD matrices, most notably through \textbf{SP-MHA}, a structure-preserving multihead attention mechanism.

In this paper, we further analyze the capabilities of SPDTransNet, and introduce a variant that allows for the inclusion of additional EEG-derived learned features within our input matrices.
We conduct a thorough examination of our model's capabilities, including its ability to perform in a multi-dataset environment, similarly to what might be encountered in clinical use.
Finally, we compare SPDTransNet with various State-of-the-Art (SOA) approaches, showcasing our model's performance and adaptability in both single-dataset and multi-dataset experiments.

Our code is available on GitHub, in the \href{https://github.com/MathieuSeraphim/SPDTransNet_plus/}{MathieuSeraphim/SPDTransNet\_plus} repository.

\section{Related Work}
\label{sec:SOA}

\subsection{Automatic Sleep Stage Scoring}
\label{ssec:SOA_scoring}

In order to determine the evolution of a subject's sleep over time, the signals originating from a PSG exam are split into fixed-length windows, or ``epochs”, with each epoch being manually ``scored” (i.e. labeled) with a corresponding sleep stage by experts.
This scoring is based on the properties of the considered signals in and around said epoch, such as frequential components and distinct punctual events.
A selection of relevant frequency bands for PSG signal analysis can be found in Table~\ref{tab:frequency_bands}.

The widespread scoring manual of the American Academy of Sleep Medicine (AASM)~\cite{berry2017aasm} defines a total of 5 sleep stages - wakefulness, REM sleep, and three stages of non-REM sleep, N1 to N3, along with the corresponding characteristics thereof, and recommends epochs of thirty seconds of length.

\begin{table}[]
\centering
\caption{Frequency bands inspired by the literature~\cite{berry2017aasm}}
\label{tab:frequency_bands}
\let\cline\clineorig
\begin{tabular}{c|c|c|c|c|c|c|}
\cline{2-7} & Delta & Theta & Alpha & Beta$_{low}$ & Beta$_{high}$ & Gamma \\ \hline
\multicolumn{1}{|c|}{Hz} & [0.5, 4[ & [4, 8[ & [8, 12[ & [12, 22[ & [22, 30[ & [30, 45[ \\ \hline
\end{tabular}
\end{table}

Most State-of-the-Art automatic sleep staging models published in recent years take sequences of PSG epochs as input, rather than individual epochs, to better account for contextual information.
Said models usually follow one of two classification schemes - sequence-to-element (sometimes called many-to-one)~\cite{chambon2017,seo2020}, or sequence-to-sequence (a.k.a. many-to-many)~\cite{supratak2017,phan2019,phan2022sleeptransformer,phan2022xsleepnet,phan2023lseqsleepnet}.
In a sequence-to-element classification scheme, the model outputs  the classification of a single epoch in the input sequence. By contrast, a sequence-to-sequence scheme outputs one classification per input epoch, and utilize a combination strategy if the input sequences overlap each other (meaning that a given epoch may be classified multiple times, each time in a different position in the sequence).
According to a recent survey by Phan et al.~\cite{Phan_survey}, sequence-to-sequence approaches tend to yield greater global performance, though this comes at the cost of methodological flexibility (Section~\ref{sec:method}).

Sequence-based classification can be done in a single step, using a Convolutional Neural Networks (CNN) to extract signal features and deliver a classification~\cite{chambon2017, phan2018}. Most notably, both Perslev et al.~\cite{perslev2019,perslev2021} and Jia et al.~\cite{jia2021SSN} treated this scoring as a segmentation problem, and utilized architectures derived from U-Net\cite{UNet}.
By contrast, Vilamala et al.~\cite{vilamala2017} and Dequidt et al.~\cite{paul} both took advantage of the aforementioned frequential characteristics of EEG signals, and analyzed EEG-derived time-frequency spectra using a standard CNN architecture pretrained for image recognition.

However, as seen in the aforementioned survey, the most common approach is to use a two-step architecture, by first extracting epoch-wise features before comparing said features in a sequence-wise submodel.
Here, CNNs are often utilized for epoch-wise feature extraction, with architectures designed for sequence analysis being utilized at the sequence level.
The DeepSleepNet model by Supratak et al.~\cite{supratak2017}, often used as a benchmark, uses a Recurrent Neural Network (RNN), as does IITNet by Seo et al.~\cite{seo2020}, which expands upon DeepSleepNet by extracting features from overlapping subwindows rather than from entire epochs.
Going one step further, Phan et al.~\cite{phan2019,phan2022xsleepnet,phan2023lseqsleepnet} and Guillot et al.~\cite{guillot2020,guillot2021} elected to use time-frequency spectra as input of their RNN-based architectures, interpreting the epoch-wise spectra as timeseries of frequency bins and analyzing them with RNNs.

With the advent of Transformer architectures~\cite{transformers}, attention mechanisms have proven to be particularly potent in the analysis of temporal sequences.
As such, these mechanisms have been applied to many fields, including sleep stage scoring.
Some of these approaches still utilize CNNs to extract intra-epoch features from 1D signals~\cite{qu2020,zhu2020,eldele2021}, while others utilize them at both epoch-wise and sequence-wise levels~\cite{Pradeepkumar2022}, with both SleepTransformer by Phan et al.~\cite{phan2022sleeptransformer} and MultiChannelSleepNet by Dai et al.~\cite{dai2023} taking time-frequency spectra as inputs.

With the exception of Dequidt et al~\cite{paul} and the pre-DeepSleepNet CNN-only approach of Chambon et al.~\cite{chambon2017}, none of the above approaches utilize large numbers of EEG electrodes, and therefore cannot take advantage of functional connectivity (Section~\ref{sec:intro}).
This is unsurprising, as the aforementioned AASM ruleset emphasizes the analysis through signal properties, and the most commonly used open sleep stage scoring datasets - SleepEDF~\cite{SleepEDF} and SHHS~\cite{SHHS1,SHHS2} - only offer two EEG signals for analysis.
That is not to say that no approach has explicitly analyzed functional connectivity between a large number of EEG signals.
For instance, Jia et al.~\cite{jia2020,jia2021} - and later Einizade et al~\cite{EINIZADE2023} - describe each epoch with a learned graph with electrodes as nodes, before comparing them through a sequence-wise spatio-temporal graph neural network (GNN). 
This approach arguably corresponds to an analysis through functional connectivity, though this encoding is likely to disregard some spatial information inherent to each electrode location, due to the permutation invariance property of GNNs.

It is in this context that we designed the SPDTransNet model, analyzing sequences of EEG epoch for the purpose of sleep stage scoring using a two-step Transformer architecture, through the lens of functional connectivity estimated through covariance matrices.
As far as we are aware, ours is the only Machine Learning approach to utilize covariance matrices for automatic sleep stage scoring.

\subsection{The Symmetric Positive Definite Manifold}
\label{ssec:SOA_manifold}

All covariance matrices are Symmetric Positive Semi-Definite (SPSD), i.e. with positive eigenvalues. Furthermore, when computed from linearly independent signals, they are fully Symmetric Positive Definite (SPD) - with strictly positive eigenvalues.
Conversely, all SPD matrices can be understood as covariance matrices between linearly independent signals.

The set of $n \times n$ SPD matrices, or $SPD(n)$, is a non-Euclidean, Riemannian (i.e. metric) manifold, meaning that applying regular Euclidean operations on such data seldom preserve its geometric structure, introducing deformations.
For instance, when interpolating between SPD matrices of same determinant using Euclidean metrics, the computed intermediate matrices tend to have a higher determinant - a phenomenon known as the ``swelling effect"~\cite{LogEuclidean}.

To counteract this, multiple Riemannian metrics have been introduced, including affine-invariant metrics~\cite{affine_invariant}.
As their name suggests, these offer interesting invariance properties, but present computational difficulties. For instance, there is no closed-form formula for the affine-invariant average of SPD matrices, and it has to be estimated.

Alternatively, LogEuclidean metrics offer similar results at a lower computational cost, though they have been shown to be less isotropic~\cite{LogEuclidean}.
They are defined using the (bijective) matrix logarithm function $log_{mat}(\cdot)$, which maps $SPD(n)$ onto its tangent space $Sym(n)$, the vector space of symmetric matrices:
\begin{equation}
log_{mat}(X) = U \cdot log(\Lambda) \cdot U^T \in Sym(n)\label{eq:logm}
\end{equation}
with $X$ $\in$ $SPD(n)$, $\Lambda$ and $U$ respectively the diagonal matrix containing the eigenvalues of $X$ and the corresponding eigenvectors, and $log(\Lambda)$ the matrix resulting from taking the element-wise natural logarithm of $\Lambda$'s diagonal elements.

More generally, this projection onto the tangent space can be parametered by a center of projection $P$ $\in$ $SPD(n)$:
\begin{equation}\label{eq:logm_P}
log_{mat}^P(X) = P^{1/2} \cdot log_{mat}(P^{-1/2} \cdot X \cdot P^{-1/2}) \cdot P^{1/2} \in Sym(n)
\end{equation}
The tangent space at point $P$ is in essence a local Euclidean approximation of the non-Euclidean manifold. As such, logarithmically mapping elements of $SPD(n)$ that are distant from the chosen parameter $P$ can introduce deformations in their mapped images.

From this, LogEuclidean metrics are defined thusly:
\begin{equation}
\label{eq:LE}
\delta_{LE}^P(X, Y) = \lVert log_{mat}(P^{-1/2} \cdot X \cdot P^{-1/2}) - log_{mat}(P^{-1/2} \cdot Y \cdot P^{-1/2}) \rVert_{*}
\end{equation}
with $X$, $Y$ and $P$ in $SPD(n)$, and $\lVert \cdot \rVert_{*}$ any Euclidean norm on $Sym(n)$.

This formula can also be written by directly using Equation~\ref{eq:logm_P}, with $\delta_{LE}^P(X, Y) = \lVert log_{mat}^P(X) - log_{mat}^P(Y) \rVert_{+}$.
Here, $\lVert S \rVert_{+}$ is equal to $\lVert P^{1/2} \cdot S \cdot P^{1/2} \rVert_{*}$ for all $S$ in $Sym(n)$~\cite{barachant2013, yger2015}.

Like their affine-invariant counterparts, LogEuclidean metrics define geodesics on the SPD manifold, and are immune to the swelling effect.
And while their relative anisotropy may affect the performance of any model utilizing them, this can be mitigated by choosing an appropriate center of projection $P$~\cite{yger2015}.
More importantly, they provide significant computational advantages, such as a closed-form formula for a Riemannian weighted sum (and by extension, for a Riemannian average):
\begin{equation}\label{eq:wFM}
\mathcal{S}um_{LE}^P(\{X\}, \{w\}) = \exp_{mat}^P\left( \sum_{i=1}^N w_i \log_{mat}^P(X_i) \right)
\end{equation}
with every $X_i$ in $\{X\}$ $\subset$ $SPD(n)$, every $w_i$ in $ \{w\}$ $\subset$ $\mathbb{R}$, and $\exp_{mat}^P(\cdot)$ the inverse of $\log_{mat}^P(\cdot)$.

\subsection{Deep Learning on the SPD Manifold}
\label{ssec:SOA_learning}

Traditional Deep Learning techniques are often implicitly designed to process Euclidean data, such as signals or images.
This limitation has led to the development of variants, to utilize these powerful techniques in a non-Euclidean setting.
Applied to the SPD manifold, Huang and Van Gool~\cite{huang2017spdnet} developed Deep Learning layers analogous to those found in CNNs, most notably their SPD-to-SPD Bilinear Mapping (BiMap) layer:
\begin{equation}
\forall X \in SPD(n),\ BiMap(X, W) = W \cdot X \cdot W^T \in SPD(m)
\end{equation}
with $W$ $\in$ $\mathbb{R}^{m \times n}$ a semi-orthogonal weights matrix.

A similar but distinct approach was taken by Chakraborty et al.~\cite{manifoldnet}, with SPD matrices being elements of a feature map, rather than the feature map itself.
In this approach, elements of $SPD(n)$ are analogous to scalars in a traditional CNN, and convolution operations are replaced by weighted Riemannian averages (Section~\ref{ssec:SOA_manifold}).

Though Riemannian methods have been utilized for a while in Brain-Computer Interfaces~\cite{BCI1,BCI2}, those have traditionally been relatively simple, such as Riemannian Minimal Distance to Mean (MDM) or SVM classifiers. Those approaches have at times been enhanced through metric learning, applied to the center of projection and/or Euclidean norm of the LogEuclidean metric~\cite{yger2015,huang2015}.
Nevertheless, Riemannian Deep Learning approaches have recently started to emerge~\cite{TANG2023,peng2023,lu2023,ouahidi2024}.

That is not to say that Deep Learning approaches are rare in BCI~\cite{BCI2}. In particular, as shown by a recent survey by Abibullaev et al.~\cite{AbibullaevSurvey}, a number of Transformer-based approaches have been applied to BCI tasks, though as far as we are aware, none take advantage of Riemannian geometry.

Beyond BCI, some Transformer-based models \textit{do} make use of Riemannian geometry.
For instance, both Konstantinidis et al.~\cite{Konstantinidis2022} and Dong et al.~\cite{DONG2023} generate SPD matrices as attention maps.
By contrast, both He et al.~\cite{he_2021} and Li et al.~\cite{li2022} adapted their Transformer architecture to handle manifold-valued data, the manifolds in question being 2D surfaces within 3D space.
More generally, Kratsios et al.~\cite{kratsios2022} developed a theoretical framework for the use of Transformers to analyze data from a variety of constrained sets, including Riemannian manifolds.

This said, we have yet to encounter a Transformer-based approach designed to analyze SPD-valued data in a structure-preserving manner, even beyond the scope of functional connectivity.

\section{Operations on SPD Matrices}
\label{sec:SPD}

Before their analysis by our model, our approach calls for a number of operations to be applied to the original $n \times n$ SPD covariance matrices - to enrich them and improve our model's performance, without sacrificing their underlying SPD structure.

\subsection{Matrix Augmentation}
\label{ssec:SPD_augmentation}

By construction, covariance matrices primarily encode comparative information, with the only monosignal information being the variance over the signal segment.
However, as hinted at by the level of performance of single-signal approaches (Section~\ref{ssec:SOA_scoring}), it is undeniable that there is relevant information contained in the signals themselves, information that may be lost when only considering covariances.
Hence, the addition of signal-wise information to our model's input should be beneficial to said model's classification performance.

Let $X$ $\in$ $SPD(n)$ be a covariance matrix encoding for the relationships between $n$ signal segments. Let $A$ $\in$ $\mathbb{R}^{n \times k}$ be a matrix encoding $k$ features for each of these signal segments, and $\alpha$ $\in$ $\mathbb{R}$ be a factor weighting for the importance of $A$ in the final matrix. These matrices may be combined to create the following SPD matrix:
\begin{equation}\label{eq:augmentation}
    X' = aug_\alpha(X, A) =
    \left(
        \begin{array}{c|c}
        \\
        X + \alpha^2 \cdot A \cdot A^T & \alpha \cdot A \\
        \\
        \hline
        \alpha \cdot A^T & \mathbb{I}_k \\
        \end{array}
    \right) \in SPD(m)
\end{equation}
with $m$ = $n + k$.

We refer to this operation as an ``augmentation", with the resulting SPD matrix $X'$ being the augmented matrix.
The model parameter $\alpha$ (called ``augmentation factor" in this paper) controls the relative influence of $A$ within the augmented matrix $X'$.

A proof of the SPD nature of matrices augmented in this manner can be found in Section~\ref{supsec:augmentation} of the Supplementary Material.

\subsection{Matrix Whitening}
\label{ssec:SPD_whitening}

When analyzing signals originating from multiple recordings, each computed covariance matrix may encode for information specific to its own recording.
As such, multi-subject datasets of biological signals, like the ones considered in this paper, can be prone to significant differences from one subject to the next.
These specificities essentially act as noise, and are liable to lower the performance of our model.

To increase comparability between recordings, we eliminate these specificities through the whitening operation~\cite{yger2015}.
Let $X'$ $\in$ $SPD(m)$ be a matrix computed from a given recording (thereafter referred to as a ``recording matrix"), and $G$ $\in$ $SPD(m)$ the matrix estimating said recording's specificities (i.e. the recording's ``whitening matrix").
The whitened matrix $M$ $\in$ $SPD(m)$ is obtained thusly:
\begin{equation}\label{eq:whitening}
M = whit(X', G) = G^{-1/2} \cdot X' \cdot G^{-1/2}
\end{equation}

This whitening can be interpreted as a transport operation.
For instance, computing a LogEuclidean operation on unwhitened matrices using $G$ as the center of projection (Equation~\ref{eq:LE}) is strictly equivalent to computing the same operation on whitened matrices, but with the identity matrix $\mathbb{I}_m$ as center of projection.

Since our matrices are computed from biological recordings of different individuals, they can encode information specific to their own recording. Therefore, within the SPD manifold, matrices corresponding to a given class and originating from a given recording can be geometrically separated from matrices corresponding to the same class but a different recording.
By choosing the proper whitening matrix $G$ for each recording, we can realign the corresponding recording matrices within a unified geometric neighborhood.

\subsection{Bijective Tokenization}
\label{ssec:SPD_tokens}

Let $\{M\}$ be a set of matrices within $SPD(m)$.
Let $M_i^*$ = $log_{mat}(M_i)$ (Equation~\ref{eq:logm}) be a bijective mapping of $M_i$ $\in$ $\{M\}$ onto $Sym(m)$, the vector space of symmetric matrices (Section~\ref{ssec:SOA_manifold}) and of dimension $d(m)$ = $\frac{m(m+1)}{2}$.
Finally, let $V_{M_i}$ be the vector representation of $M^*_i$ within the vector space $\mathbb{R}^{d(m)}$, obtained by describing $M_i^*$ in the canonical basis of $Sym(m)$.

It follows that any weighted sum or linear combination between vectors in $\{V_{M}\}$ $\subset$ $\mathbb{R}^{d(m)}$ is equivalent to the same operation between the corresponding symmetric matrices in $\{M^*\}$, itself equivalent to a LogEuclidean (and therefore Riemannian) operation between the corresponding SPD matrices within $\{M\}$ (Equation~\ref{eq:wFM}), parametered by $P$ equal to the identity matrix $\mathbb{I}_m$.

Therefore, by bijectively mapping matrices in $SPD(m)$ onto $\mathbb{R}^{d(m)}$, we can apply some Riemannian operations between our matrices through Euclidean operations in $\mathbb{R}^{d(m)}$.
Equivalently, given any triangular number $k$ (so that $k$ = $\frac{j(j+1)}{2}$), any Euclidean weighted sum between elements of $\mathbb{R}^{k}$ is equivalent to a Riemannian operation within $SPD(j)$.

We refer to this mapping between $SPD(m)$ and $\mathbb{R}^{d(m)}$ as the ``\textit{tokenization}" operation\footnote{Following the common convention of referring to the vector inputs of Transformer-based architectures as ``tokens".}, and to $\mathbb{R}^{d(m)}$ as a ``\textit{triangular vector space}".

\subsection{Triangular Maps}
\label{ssec:SPD_maps}

Let $\mathcal{L}_{m, p}(\cdot)$ be a linear map between $\mathbb{R}^{d(m)}$ and $\mathbb{R}^{d(p)}$. As stated above, it can also be understood as a linear map between $Sym(m)$ and $Sym(p)$, with the sets' vector representations as intermediaries.

In~\cite{arxiv}, we used $\mathcal{L}_{m, p}(\cdot)$ to define a mapping between $SPD(m)$ and $SPD(p)$:
\begin{equation}\label{eq:SPD_mapping}
\mathcal{L}^\mathcal{R}_{m, p}(M) = exp_{mat} \circ \mathcal{L}_{m, p} \circ log_{mat}(M) \in SPD(p)
\end{equation}
with $M$ $\in$ $SPD(m)$, $log_{mat}(\cdot)$ the matrix logarithm and $exp_{mat}(\cdot)$ the matrix exponential (Section~\ref{ssec:SOA_manifold})\footnote{Here, the logarithm is defined from $SPD(m)$ to $Sym(m)$, and the exponential from $Sym(p)$ to $SPD(p)$.}.

By this definition, applying $\mathcal{L}_{m, p}(\cdot)$ on vectors in $\mathbb{R}^{d(m)}$ is equivalent to applying $\mathcal{L}^\mathcal{R}_{m, p}(\cdot)$ on matrices in $SPD(m)$.

Since $\mathcal{L}_{m, p}(\cdot)$ is a continuous linear map, it preserves relations of proximity:
\begin{equation}\label{eq:ineq_1}
\lVert \mathcal{L}_{m, p}(V_A) - \mathcal{L}_{m, p}(V_B) \rVert_2 \leq \lVert W \rVert_\bullet \cdot \lVert V_A - V_B \rVert_2
\end{equation}
with $V_A, V_B$ $\in$ $\mathbb{R}^{d(m)}$, $W$ $\in$ $\mathbb{R}^{d(p) \times d(m)}$ the transformation matrix of $\mathcal{L}_{m, p}(\cdot)$, $\lVert \cdot \rVert_2$ the $L^2$ norm for vector representations, and $\lVert \cdot \rVert_\bullet$ the matrix norm induced by $\lVert \cdot \rVert_2$.

By the definition of LogEuclidean metrics (Equation~\ref{eq:LE}), and with $A, B$ $\in$ $SPD(m)$ the matrices that tokenize into $V_A$ and $V_B$, we can rewrite Equation~\ref{eq:ineq_1} thusly:
\begin{equation}\label{eq:ineq_2}
\delta_{LE}^{\mathbb{I}_p}(\mathcal{L}^\mathcal{R}_{m, p}(A), \mathcal{L}^\mathcal{R}_{m, p}(B)) \leq \lVert W \rVert_\bullet \cdot \delta_{LE}^{\mathbb{I}_m}(A, B)
\end{equation}
Hence, the SPD-to-SPD mapping $\mathcal{L}^\mathcal{R}_{m, p}(\cdot)$ preserves proximity between elements of the source manifold - and by extension, the underlying SPD structure of the input data.

Note that unlike the BiMap SPD-to-SPD linear mapping (Section~\ref{ssec:SOA_learning}), our $\mathcal{L}^\mathcal{R}_{m, p}(\cdot)$ mapping does not require restrictions to be imposed on the weights matrix $W$.

\section{The SPDTransNet Model}
\label{sec:model}

The SPDTransNet model, first introduced in~\cite{arxiv}, can be seen Figure~\ref{fig:model}. As stated in Section~\ref{sec:intro}, it expends upon an earlier iteration of our approach~\cite{CAIP}, itself based on the SleepTransformer architecture of Phan et al.~\cite{phan2022sleeptransformer}, and follows a Transformer-based~\cite{transformers} architecture, as those are particularly well suited to the analysis of vector sequences.

To better account for contextual information, as with much of the SOA approaches (Section~\ref{ssec:SOA_scoring}), our model takes a sequence of epochs as input, and follows a two-step architecture: an intra-epoch step to extract epoch-wise features, followed by an inter-epoch (or sequence-wise) step to compare them before the final classification.
In particular, SPDTransNet takes a sequence of length $L$ = $2 \cdot \ell + 1$, composed of a central epoch flanked by its $\ell$ preceding and following epochs (Figure~\ref{fig:model}).

\subsection{Data Preparation}
\label{ssec:model_data}

\begin{figure}[h]
\centering
\includegraphics[width=1\textwidth]{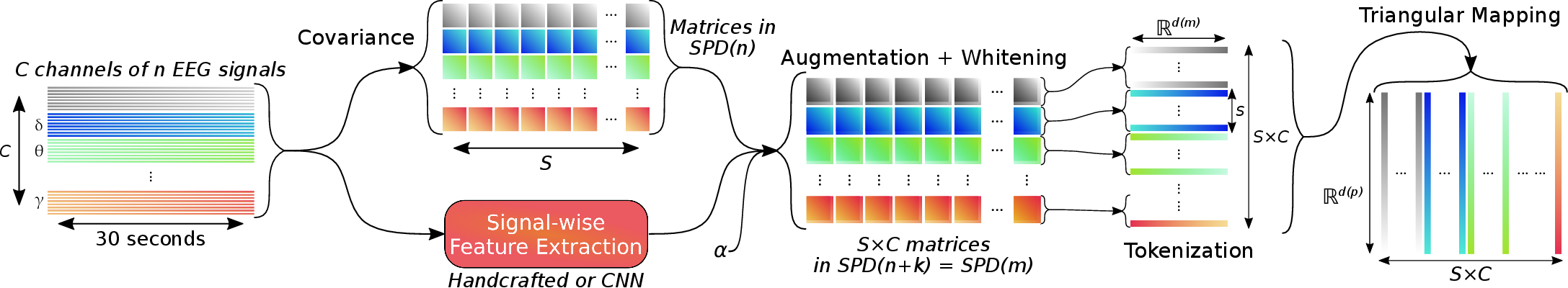}
\caption{Our data preparation pipeline, with $S$ = 30, $C$ = 7 and $n$ = 8.}
\label{fig:signals_to_tokens}
\end{figure}

Using the concepts developed in Section~\ref{sec:SPD}, for each EEG epoch, we process a collection of signals into a sequence of tokens with an underlying SPD structure, as seen in Figure~\ref{fig:signals_to_tokens}.

\subsubsection{Building SPD matrices}
\label{sssec:model_data_covariances}

After selecting $n$ signals to compose our matrices, our first step is to apply a z-score normalization on each signal, to ensure comparability between all signals, both between recordings and within each recording.
Additionally, for each of the six frequency bands defined in Table~\ref{tab:frequency_bands}, we apply a fourth order Butterworth bandpass filter, to isolate different frequential components of the signal.
This results in a total of $C$ = 7 input channels, six derived from filtered signals and one from the unfiltered signal, corresponding to the input in Figure~\ref{fig:signals_to_tokens}.

For each channel, the signals are subdivided into fixed-length segments, without overlap between segments.
We chose to fix  the segments' length to one second, as this roughly corresponds to the duration of events indicative of a given sleep stage within the signal~\cite{berry2017aasm}.
We then compute an $n \times n$ covariance matrix for each segment of each channel, netting us a grid of $30 \times 7$ such matrices per epoch.

\subsubsection{Enriching our Matrices}
\label{sssec:model_data_enrichment}

Once computed, the matrices are then enriched through augmentation and whitening  (Sections~\ref{ssec:SPD_augmentation} and~\ref{ssec:SPD_whitening}), with distinct whitening matrices computed for each channel (Figure~\ref{fig:signals_to_tokens}).

\paragraph{Augmentation Strategies}
\label{ssssec:model_data_enrichment_augmentation}

Let $\{X\}$ $\subset$ $SPD(n)$ be the set of all covariance matrices for a given channel of a given recording.
As stated in Section~\ref{ssec:SPD_augmentation}, the goal of the augmentation operation is to add signal-specific information to each covariance matrix $X_i$ $\in$ $\{X\}$, through the augmentation matrix $A_i$ $\in$ $\mathbb{R}^{n \times k}$ weighted by $\alpha$ $\in$ $\mathbb{R}$.
With $X_i$ a constant, and for a given value of $\alpha$, the output of the augmentation operation is entirely determined by the matrix $A_i$.
To compute it, we have developed two strategies.

Firstly, we can compute $A_i$ through signal-wise handcrafted features. That is to say, for each of the $n$ signal segments used to construct $X_i$, we compute $k$ segment-wise statistics.
In this paper, we utilize the average power spectral density (PSD) as the relevant statistic, as we have found it to be the best performing amongst a number of tested statistics in our previous work~\cite{CAIP}.
Having tested the use of a combination of statistics to construct $A_i$ (i.e. $k > 1$), we have found no improvement in performance, and have therefore elected to only use the average PSD of each signal as handcrafted features (keeping $k$ = 1).

Our second strategy is to learn the augmentation matrix $A_i$ using a dedicated pretrained submodel, integrating said submodel within our architecture and finetuning it during training.
As a first iteration, we have chosen to utilize the epoch-wise feature extraction CNN developed by Seo et al. for IITNet~\cite{seo2020}, as it is designed to analyze subwindows of each input epoch, delivering a local feature vector for each subwindow.
We have modified it to take our $n$ signals over $C$ channels per input (Section~\ref{sssec:model_data_covariances}).
The CNN's weights are shared between the $n$ signals in a given channel, but not between channels, since our channel-wise filtering necessarily changes the signals' spectral configuration between channels - and therefore the required CNN embedding.
More details about the learned augmentation submodel can be found in Section~\ref{supsec:IITNet} of the Supplementary Material.

When utilizing this strategy, in order to not drown the covariance information within the augmented matrix, we impose $k$ = 3.

\paragraph{Composition with Whitening}
\label{ssssec:model_data_enrichment_whitening}

As stated in Section~\ref{ssec:SPD_whitening}, the whitening operation can be used to transport matrices computed from different recordings onto the same neighborhood within the SPD manifold. As such, we  have elected to operate the whitening operation after the augmentation, regardless of whichever augmentation strategy is chosen.

Let $I$ denote the set of epochs for a given recording and channel, and let $\{X_i\}_{i\in I}$ and $\{A_i\}_{i\in I}$ be the computed covariance matrices and their augmentation matrices, respectively.
A first definition of their whitening matrix is as follows:
\begin{equation}\label{eq:DAW}
G = \mathcal{A}vg_{AI}\left(\{aug_\alpha(X_i, A_i)\}_{i\in I}\right)
\end{equation}
where $\mathcal{A}vg_{AI}(\cdot,\cdot)$ denotes the standard affine-invariant average\footnote{As implemented in~\cite{pyriemann}.} (Section~\ref{ssec:SOA_manifold}), and $aug_\alpha(\cdot,\cdot)$ is the augmentation operator (Equation~\ref{eq:augmentation}).
We refer to this composition strategy as ``direct average whitening" (DAW).

Whitening operations performed through Equations~\ref{eq:whitening} and~\ref{eq:DAW} shift all the relevant augmented matrices around the identity matrix $\mathbb{I}_m$. In other words, the affine-invariant average of all matrices enriched through DAW is $\mathbb{I}_m$.
While this shift has been shown to improve performance in some contexts~\cite{barachant2013}, the introduction of augmentation raises some issues.
Namely, that Equation~\ref{eq:DAW} computes in a single step an average value derived from both matrices in $\{X_i\}_{i\in I}$  and in $\{A_i\}_{i\in I}$. 
While $\{X_i\}_{i\in I}$ is composed of SPD matrices, and is therefore nicely resumed by an affine-invariant average, $\{A_i\}_{i\in I}$ is composed of matrices of attributes, and would be better represented by an Euclidean mean.

Based on this, we have proposed two different alternative scheme to compute each whitening matrix, starting with our ``mirrored augmentation whitening" (MAW) composition strategy, which defines $G$ thusly:
\begin{equation}\label{eq:MAW}
\begin{gathered}
    G' = \mathcal{A}vg_{AI}(\{X_i\}_{i\in I}) \;;\;   A_G' = \mathcal{A}vg_E(\{A_i\}_{i\in I}) \\
    G = aug_\alpha(G', A_G')
\end{gathered}
\end{equation}
with $\mathcal{A}vg_E(\cdot)$ the standard Euclidean average.

Both DAW and MAW configurations augment their matrices prior to whitening. For the sake of comparison, we elected to test the reverse configuration.
We define the resulting ``whitening prior to augmentation" (WPA) thusly:
\begin{equation}\label{eq:WPA}
\begin{gathered}
    G' = \mathcal{A}vg_{AI}(\{X_i\}_{i\in I}) \;;\; \{X^*_i\}_{i\in I} = whit(\{X_i\}_{i\in I}, G') \\
    \{M_i\}_{i\in I} = aug_\alpha(\{X^*_i\}_{i\in I}, \{A_i\}_{i\in I})
\end{gathered}
\end{equation}

\begin{figure}[h]
\centering
\includegraphics[width=1\textwidth]{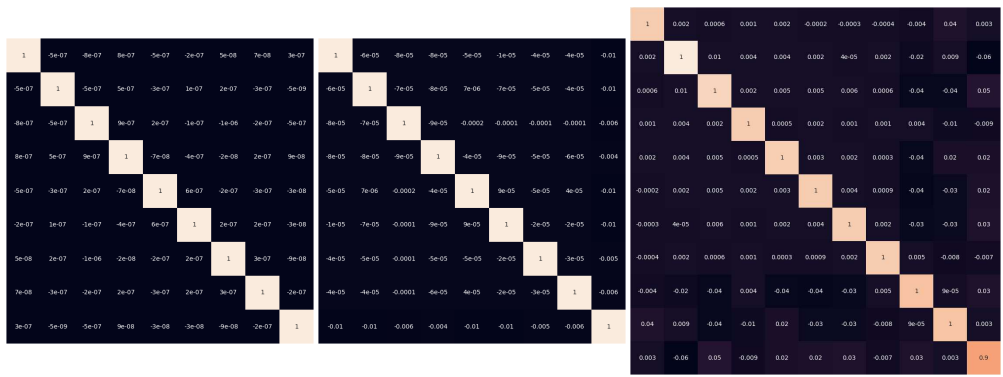}
\caption{Estimation of the affine-invariant average of recording-wise enriched matrices in heatmap form, computed from unfiltered signals on the MASS-SS3 recording of index 42. From left to right, these correspond to DAW, MAW and WPA enrichment, respectively.}
\label{fig:enrichment_averages}
\end{figure}

Shown in Figure~\ref{fig:enrichment_averages} are subject-wise average matrices resulting from all three enrichment strategies, as obtained for each strategy's highest performing configuration (entries 1, 2 and 4 in Table~\ref{tab:ablation_enrichment}). As such, the displayed WPA-enriched average matrix utilizes learned augmentation features, while the other two utilize handcrafted ones.

As we can see in the figure, although the DAW strategy leads to the closest augmented matrix to the identity, with no non-diagonal element greater than $10^{-6}$ in absolute value, the MAW and WPA strategy still leads to matrices relatively close to said identity.

As this closeness to the identity remains true for other recordings and channels\footnote{As can be seen in the equivalent matrices computed for each recording and channel, available in our GitHub repository.}, we may empirically conclude that all three composition strategies transport our matrices upon the same neighborhood within the SPD manifold.

\subsubsection{Final Input Processing}
\label{sssec:model_data_input}

Following the matrix enrichment operations, from the original covariance matrices in $SPD(n)$, we obtain matrices in $SPD(m)$ ($m \geq n$).
These are then tokenized (Section~\ref{ssec:SPD_tokens}), yielding a total of $30 \times 7$ tokens of $\mathbb{R}^{d(m)}$ per epoch.
In order to combine all channels into a single dimension, as seen in Figure~\ref{fig:signals_to_tokens}, the tokens are rearranged into a single sequence of length 210, with the thirty tokens of channel 0 being followed by the thirty tokens of channel 1, etc.

As we have found that relatively small token sizes lead to worse performance\footnote{Given a token size $d$ and a number of attention heads $h$ for a Transformer encoder, we found that imposing $\frac{d}{h} \geq 32$ yielded better results.}, we linearly map our tokens (Section~\ref{ssec:SPD_maps}) onto a larger dimension $\mathbb{R}^{d(p)}$ ($p > m$) before going forward to the intra-epoch component.

Note that the most time-consuming operation in our training pipeline is the eigenvalue decomposition necessary to compute the matrix logarithm, itself a component of our bijective tokenization (Section~\ref{ssec:SPD_tokens}).
We implement this operation through a singular value decomposition (SVD) algorithm. More precisely, for the purposes of backpropagation, we approximate the SVD operator's gradient using Taylor polynomials, as this computation tends to be numerically unstable otherwise~\cite{svd}.

In order to speed up model training, whenever the enrichment process is invariant ( i.e. the augmentation factor $\alpha$ is non-trainable and handcrafted features are utilized, cf. Section~\ref{ssssec:model_data_enrichment_augmentation}), we tokenize our matrices prior to the start of the learning process, rather than at each training step, which results in a significant speed-up.

\subsection{Model Architecture}
\label{ssec:model_architecture}

\begin{figure}[h]
\centering
\includegraphics[width=1\textwidth]{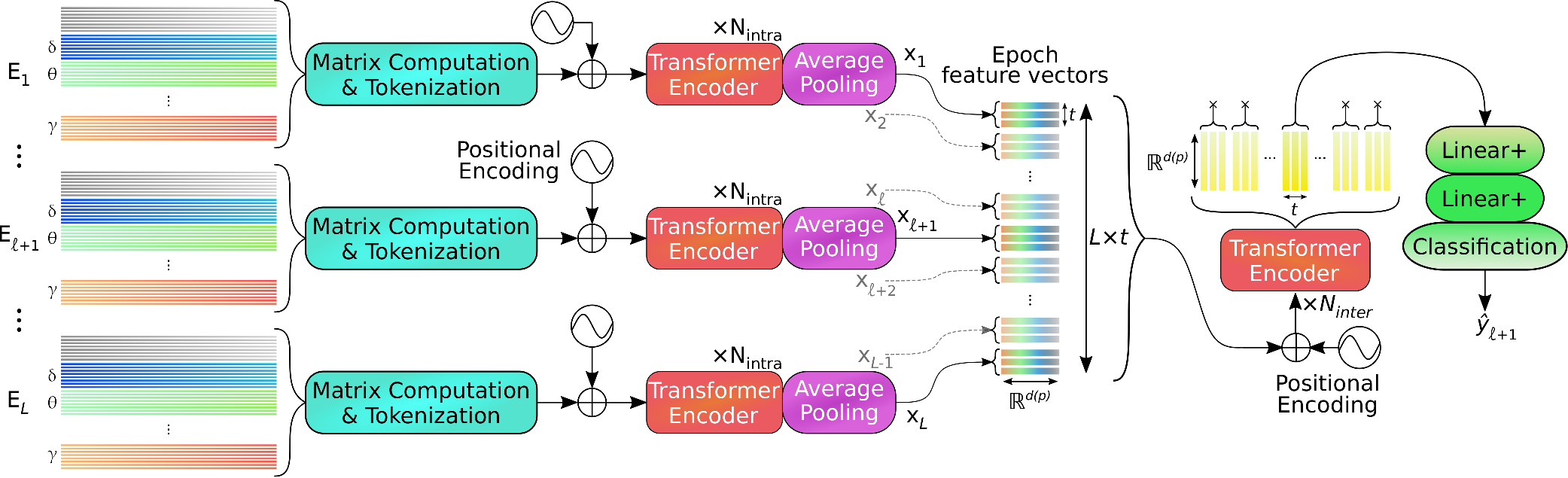}
\caption{Architecture of SPDTransNet, with $t$ = 3 feature tokens per epoch, $L$ = $2 \cdot \ell + 1$ the length of the input epoch sequence, and $\ell + 1$ the index of the central epoch. The ``Matrix Computation \& Tokenization" component is further developed in Figure~\ref{fig:signals_to_tokens}.}
\label{fig:model}
\end{figure}

As presented in Figure~\ref{fig:model}, after a learnable positional encoding~\cite{posenc}, our tokens pass trough an intra-epoch Transformer encoder, whose output sequence is then averaged into $t$ epoch feature tokens.
While only $t$ = $1$ (single feature token) and $t$ = $7$ (one feature token per input channel) have interpretations connecting the output of the average pooling layer and the input of the encoder, we tested other values of $t$ for the sake of completeness.

The epoch-wise features are then combined into a $L$ $\times$ $t$ sequence, and compared through the inter-epoch Transformer encoder, after which the $t$ tokens corresponding to the central epoch pass through two linear (i.e. fully connected) layers with dropout and ReLU activation (labeled ``Linear+" in Figure~\ref{fig:model}), followed by a third and final linear layer outputting the classification vector.
Finally, the training performance is ascertained through a cross-entropy loss function (with label smoothing~\cite{label_smoothing} for additional regularization), complete with application of a logarithmic softmax function to the classification vector.

As such, our model follows a sequence-to-element classification scheme (for reasons explained in Section~\ref{sec:method}), but can easily be adapted to follow a sequence-to-sequence scheme instead (Section~\ref{ssec:SOA_scoring}).

\subsection{Model Structural Preservation}
\label{ssec:model_struct}

As seen in Section~\ref{ssec:SPD_maps}, weighted sums and triangular linear maps involving tokenized matrices don't cause issue when it comes to SPD structural preservation.
Hence, standard fully connected / linear neural network layers are permissible, as long as the output length is triangular. Furthermore, the addition of a bias term doesn't cause any additional issue, as it boils down to a weighted sum between vector representations.

This said, our SPDTransNet model contains a number of other operations, whose structure preservation properties need to be further justified.

\subsubsection{Structure-Preserving Multihead Attention (SP-MHA)}
\label{sssec:model_struct_mha}

\begin{figure}[h]
\centering
\includegraphics[width=.4\textwidth]{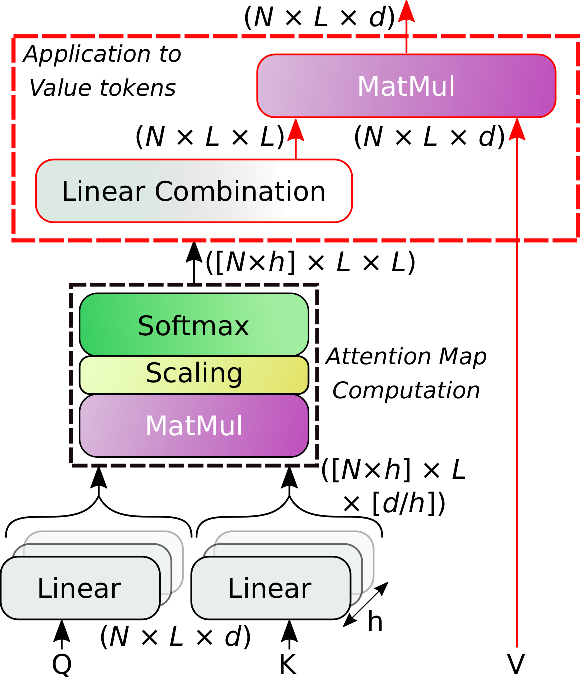}
\caption{Our SP-MHA component. The computation of attention maps (small-dashed black rectangles) is identical to the original MHA, whereas the application of said attention maps to the tokens within the \textbf{V}alue tensor (large-dashed red rectangle) has been modified to avoid any projection and subsequent concatenation.}
\label{fig:mha}
\end{figure}

Classically, Transformers utilize a component called Linear Multihead Attention (L-MHA), first introduced by Vaswani et al.~\cite{transformers}
Unfortunately, this component calls for the concatenation of output tokens for each attention head, an operation that is not interpretable using our framework of tokenized SPD matrices.

To counteract this, in our previous work~\cite{arxiv}, we defined a structure-preserving alternative to L-MHA, called SP-MHA (as seen in Figure~\ref{fig:mha}).
We proved that not only did our SP-MHA preserve the structure of our input tokens, but it did so without requiring linear maps, aside for the attention map computation (the small-dashed black rectangle in the figure) - which remains identical to L-MHA.

\subsubsection{Additional Guarantees}
\label{sssec:model_struct_extra}

Aside from the final classification layer and the computation of attention maps (Section~\ref{sssec:model_struct_mha}), all linear mappings of our tokens are triangular (Section~\ref{ssec:SPD_maps}).
Notably, this includes the two mappings contained in the Feed-Forward (FF) component of Transformer encoders~\cite{transformers}.

As sums and linear combinations of tokens are permissible, most other model components - i.e. positional encodings, average poolings and in-encoder layer normalizations - do not cause a loss of structure.
Same for the ReLU and dropout layers in the Transformer encoders' FF components, as setting values within a token to zero won't remove the corresponding matrix from $Sym(m)$.
The only issue remaining is the first Linear+ layer, which combines the $t$ feature tokens output by the inter-epoch encoder by flattening them - hence losing the interpretability of its output as SPD matrices.
However, as these layers are intended to translate the embedding within the Transformer encoders into the final classification, rather than being part of said embedding themselves, this doesn't cause issue.

Therefore, we claim structure preservation for the data going through the SPDTransNet model up to the final classification layer for $t$ = 1, and up to the first Linear+ layer for $t$ $>$ 1.

\section{Experimental Methodology}
\label{sec:method}

Due to the high degree of imbalance within sleep stage scoring datasets, along with low performance of SOA methods for some sleep stages (Section~\ref{sec:comparison}), we have elected to orient our methodology towards maximizing per-stage performance.
As such, we use the macro-averaged (i.e. unweighted average of a binary statistic computed for each class) F1 score, or MF1, as main performance metric, as it is insensitive to class imbalance and is often used in the literature in this context.

In addition, to ensure that the influence of the least represented classes isn't eclipsed during the learning process, we rebalance our training sets through oversampling - resulting in the same number of examples for each class. This requires the assignment of a unique label to each input; as such, we cannot use a (multi-label) sequence-to-sequence classification scheme (Section~\ref{ssec:SOA_scoring}).
For each configuration tested, we evaluate our model on a given dataset through cross-validation, preceded by a hyperparameter research on a single fold.

In Section~\ref{sec:comparison}, we compare our model to a number of sequence-based State-of-the-Art approaches.
However, the corresponding models do not share our input sequence definition.
In order to counteract border effects and ensure that the same epochs are classified in all models within the reported results, we ignore some epochs at the beginning and end of each test set recording, as done in our previous work~\cite{CAIP,arxiv}.
More precisely, we ensure that the first and last 24 epochs are ignored\footnote{The largest possible number of epochs ignored by DeepSleepNet~\cite{supratak2017} at the end of a test set recording.}. For SPDTransNet, it means that the first and last 24 - $\ell$ epochs (Section~\ref{sec:model}) in each test set recording will not ever be used, even as context epochs.
We refer to this as ``test set clipping" in subsequent sections.

\subsection{SPDTransNet Configuration}
\label{ssec:method_config}

As with SleepTransformer~\cite{phan2022sleeptransformer}, we have chosen a default input sequence length of $L$ = 21 for SPDTransNet, corresponding to a context size of $\ell$ = 10 (Section~\ref{sec:model}).
To avoid cluttering the hyperparameter research process, we chose constant values for some low-impact hyperparameters, for instance setting all dropout rates to 0.1, and the hidden dimension of our Transformer encoders' feedforward components~\cite{transformers} to $d(x)$ = 903 (i.e. $x$ = 42, cf. Section~\ref{ssec:SPD_tokens}).

For the hyperparameter research (Section~\ref{sec:method}), we utilize the Optuna tool~\cite{optuna}, with five simultaneous runs and 50 total runs per configuration (or eight simultaneous runs if the augmentation features are learned, as this significantly lengthens the learning process - see thereafter).
Through it, the token size $d(p)$ following the first triangular map (Section~\ref{sssec:model_data_input}) is chosen in \{351, 378\} (i.e. $p$ $\in$ \{26, 27\}), and the $h$ parameter of each Transformer encoder (Section~\ref{sssec:model_struct_mha}) in \{3, 9\}  - a restricted choice, since $\frac{d(p)}{h}$ must remain an integer whatever the chosen combination.
In turn, the number of epoch feature tokens $t$ (Section~\ref{ssec:model_architecture}) among \{1, 3, 5, 7, 10\}.
Other parameters, like the learning rate and augmentation factor $\alpha$ (Sections~\ref{ssec:SPD_augmentation} and~\ref{sssec:model_data_enrichment}), are chosen from a range using log-uniform sampling.

In our testing, we have found that moderate variations in the factor $\alpha$ have a small impact on classification when the network is configured for augmentation through handcrafted features (Section~\ref{ssssec:model_data_enrichment_augmentation}).
As such, when allowed to be learned, $\alpha$ remains quasi-constant.
To avoid unnecessary computations at each step, in this configuration, we set $\alpha$ as constant, enabling us to speed up training times by pre-tokenizing our matrices (Section~\ref{sssec:model_data_input}).
This is not the case when utilizing learned features for augmentation, and $\alpha$ is kept learnable in this context.

More details can be found in our GitHub repository.

\subsection{Datasets used}
\label{ssec:method_datasets}

To best compare ourselves to the literature, we selected publicly available sleep stage scoring datasets that had a relatively large and varied number of EEG signals to choose from, since our model is based on comparison of brain activity between different regions.

\begin{table}[]
\centering
\caption{Overview of the tested datasets. Subject average age shown with standard deviation.}
\label{tab:data}
\begin{tabular}{|c|c|c|c|c|c|c|c|}
\hline
Dataset & Avg. age (yrs) & \# epochs & \% N3 & \% N2 & \% N1 & \% REM & \% Awake \\ \hline
MASS-SS1 & 63.6 $\pm$ 5.3 & 51292 & 6.64 & 43.22 & 13.87 & 12.41 & 23.87 \\ \hline
MASS-SS3 & 42.5 $\pm$ 18.9 & 59317 & 12.90 & 50.24 & 8.16 & 17.84 & 10.86 \\ \hline
Dreem DOD-H & 35.32 $\pm$ 7.51 & 24662 & 14.25 & 48.17 & 6.10 & 19.17 & 12.31 \\ \hline
\end{tabular}
\end{table}

The Montreal Archive of Sleep Studies (MASS)~\cite{MASS} is a dataset composed of five subsets (SS1 to SS5), containing a large number of common EEG signals (16 common signals in all but MASS-SS4) acquired with a common reference electrode\footnote{All MASS-SS3 subjects use a linked-ear reference (LER), as do some MASS-SS1 subjects, the others using a computed linked-ear (CLE) reference.}.
Of these, we selected eight signals corresponding to electrodes F3, F4, C3, C4, T3, T4, O1 and O2 - i.e. frontal, central, temporal and occipital locations on both hemispheres, giving us a reasonably wide range of sampled localized cerebral activity.
We selected for analysis only the subsets scored with the AASM ruleset:

\textbf{MASS-SS3} is the largest and most widely used of the tested datasets, with 62 healthy subjects.
With it, we use the seemingly random 31 folds generated by Seo et al.~\cite{seo2020}, with each fold having a 50/10/2 division of subjects dedicated to the training, validation and test sets respectively.
This follows a leave-two-out cross-validation scheme, with the union of all test sets corresponding to the dataset in its entirety.

\textbf{MASS-SS1} is made of 53 older subjects (Table~\ref{tab:data}), including 15 that had been diagnosed with mild cognitive impairment (MCI), making it harder to classify. 
We randomly generated 26 cross-validation folds, with a 42/9/2 split for all but one fold, said fold having a 42/8/3 split instead. As with MASS-SS3, we ensure that the union of test sets corresponds to the entire dataset.

Finally, we also tested our approach on the Dreem dataset~\cite{guillot2020}, an openly available dataset composed of subjects both healthy and suffering from sleep apnea. The recordings in its \textbf{Dreem DOD-H} subset, corresponding to all healthy subjects, contain 12 EEG-derived signals, including five obtained from an EEG electrode and a reference\footnote{Located behind the ear in the opposite hemisphere.}: F3, F4, C3, Fp1 and Fp2 - the latter two being pre-frontal. The remaining derivations are acquired between pairs of EEG electrodes, but we can recover the signals corresponding to O1 and O2 from them.
We have elected to only utilize these seven EEG signals in this paper, as we wish to compute our matrices (and estimate functional connectivity) between signals corresponding only to localized brain activity.
As with MASS-SS1, we have randomly generated our folds, this time 25 folds following a leave-one-out scheme and a 20/4/1 split.

As shown in Table~\ref{tab:data}, our training sets contain at most 59k epochs, without accounting for border effects or test set clipping (Section~\ref{sec:method}).
This is a relatively small dataset for training a Transformer-based model of this size, a limitation due to the relative scarcity of publicly available sleep stage scoring datasets having a sufficient amount of EEG electrodes included.
As a consequence, the cross-validation folds of each dataset are relatively small (see thereafter). Combined with a high level of inter-subject (and thus inter-recording) variability due to the biological nature of our data, fold-wise results tend to differ significantly, resulting in inflated values of standard deviation.

For the purposes of hyperparameter researches (Section~\ref{ssec:method_config}), when training on MASS-SS3, we use the same randomly selected fold as our previous work~\cite{CAIP,arxiv}.
For each of the other two datasets, a new fold was randomly selected for this purpose.

\section{Ablation Study}
\label{sec:ablation}

In order to better understand the impact of SPDTransNet's components and configurations the model's performance, we evaluated and compared a number of model variations.
These experiments utilize exclusively the MASS-SS3 dataset (Section~\ref{ssec:method_datasets}), due to its size and widespread use.

For all tables introduced in this section and Section~\ref{sec:comparison}, the term after the $\pm$ symbol corresponds to the standard deviation between the results obtained from each relevant cross-validation fold.
As stated in Section~\ref{ssec:method_datasets}, this value is somewhat inflated, and doesn't exactly correspond to an uncertainty value.
In addition, only the class-wise F1 scores for sleep stages N1 to N3 are displayed, as those are particularly relevant - with N2 being the easiest to classify, showcasing peak class-wise performance, and N1 and/or N3 being the worst performing.
Full results are available on our GitHub repository.

\subsection{Enrichment configurations}
\label{ssec:ablation_enrichment}

As presented in Section~\ref{sssec:model_data_enrichment}, prior to tokenization, our SPD matrices are enriched through a combination of augmentation and whitening, adding additional information to said matrices while ensuring their comparability by transporting matrices from different recordings and channels onto the same neighborhood.

As we can see in Table~\ref{tab:ablation_enrichment}, when using handcrafted augmentation features, the MAW strategy is the best performing, though WPA isn't significantly behind.
By contrast, the DAW strategy performed poorly, particularly in the N1 sleep stage. This confirms that the single-step averaging of combined SPD covariance matrices and Euclidean augmentation matrices would lead to data degradation, hurting the model's performance.

Due to the learning process shifting the weights used to compute learned augmentation features at every learning step, utilizing the DAW strategy in a learned augmentation configuration would add a layer of complexity to the already inflated cost of this approach (Section~\ref{ssec:method_config}).
Given this fact, compounded with the low performance differential between MAW and WPA, we have elected to only test the learned augmentation configuration with the WPA strategy.

As seen in the table, this results in a performance improvement of .4 points - relatively minimal when compared to the standard deviation of both variants.
Since the above experiments do not alter the main, post-tokenization structure-preserving model, we have elected to utilize the handcrafted augmentation and MAW strategy as our baseline enrichment configuration in Section~\ref{ssec:ablation_general}.

\begin{table}[]
\centering
\caption{Comparison of enrichment configurations for SPDTransNet, using MASS-SS3}
\label{tab:ablation_enrichment}
\begin{tabular}{|c|c|c|c|c|c|c|}
\hline
Augmentation & Whitening & MF1 & N3 F1 & N2 F1 & N1 F1 \\ \hline
Handcrafted & DAW & 79.66 $\pm$ 3.68 & 78.44 $\pm$ 10.93 & 87.77 $\pm$ 2.42 & 56.47 $\pm$ 8.74 \\ \hline
Handcrafted & MAW & 81.24 $\pm$ 3.29 & 78.81 $\pm$ 11.17 & 89.36 $\pm$ 2.36 & \textbf{60.50} $\pm$ 6.18 \\ \hline
Handcrafted & WPA & 80.97 $\pm$ 3.06 & 79.79 $\pm$ 10.27 & \textbf{89.73} $\pm$ 2.04 & 59.21 $\pm$ 7.06 \\ \hline
Learned & WPA & \textbf{81.64} $\pm$ 2.88 & \textbf{80.15} $\pm$ 10.46 & 89.54 $\pm$ 2.46 & 60.29 $\pm$ 5.52 \\ \hline
\end{tabular}
\end{table}

\subsection{Further Ablation Experiments}
\label{ssec:ablation_general}

Starting with the baseline configuration of SPDTransNet (Section~\ref{ssec:ablation_enrichment}), we further analyze our model through punctual, non-cumulative modifications to its structure and to our methodology, with results presented in Table~\ref{tab:ablation_general}.

Though the different composition strategies for augmentation and whitening were investigated in Section~\ref{ssec:ablation_enrichment}, their individual contribution to classification was not.
In the case of handcrafted augmentation features, we know that variations in the factor $\alpha$ have a negligible effect on said performance (Section~\ref{ssec:method_config}), putting into question the augmentation's influence on our classification's performance.
To learn more, we modified the baseline configuration to augment our matrices with vectors of zeros.
As seen in the table, this \textbf{zero-valued augmentation} leads to a moderate performance drop, though one slightly larger than when using the WPA strategy (Table~\ref{tab:ablation_enrichment}).
Hence, the addition of handcrafted features through augmentation \textit{does} result in an increase in performance.

In an earlier publication~\cite{CAIP}, we computed the MAW whitening matrices $G'$ (Equation~\ref{eq:MAW}) by taking the \textbf{global covariance matrix} of each recording and channel.
This is equivalent to the Euclidean average of said recording's one-second matrices, and therefore ill-suited to an analysis based on Riemannian structural preservation, as evidenced by the performance drop seen in Table~\ref{tab:ablation_general} - hence our decision to utilize the affine-invariant average instead.

A major component of SPDTransNet is the replacement of our Transformer encoders' standard L-MHA component with our own SP-MHA (Section~\ref{sssec:model_struct_mha}). To better understand its contribution to our classification accuracy, we tested our model's performance when equipped with the more \textbf{classic MHA}.
As seen in Table~\ref{tab:ablation_general}, this change leads to a slight drop in performance.
Hence, though our lead when using SP-MHA is not significant, it does mean that this component does not underperform when compared to L-MHA.
As such, the added interpretability afforded by our model's structure-preserving operations throughout the analysis (Section~\ref{sssec:model_struct_extra}) comes at no additional performance cost.

Finally, we have elected to study the input sequence length $L$ (Section~\ref{sec:model}) on SPDTransNet.
As stated in Section~\ref{ssec:method_config}, our initial choice of context size (i.e. $\ell$ = 10) is semi-arbitrary. Hence, we implement a \textbf{change in input length}, testing both $L$ = 13 (i.e. $\ell$ = 6) and $L$ = 29 (i.e. $\ell$ = 14).
However, as seen in the table, both alterations yield a reduction in performance, though the negative impact of increasing contextual information seems to outweigh that of reducing it. The baseline value of $\ell$ = 10 therefore seems to be a good compromise.

\begin{table}[]
\centering
\caption{Analysis of various modifications to the SPDTransNet baseline, using MASS-SS3}
\label{tab:ablation_general}
\begin{tabular}{|c|c|c|c|c|c|}
\hline
Configuration & MF1 & N3 F1 & N2 F1 & N1 F1 \\ \hline
\textbf{Baseline} & \textbf{81.24} $\pm$ 3.29 & \textbf{78.81} $\pm$ 11.17 & \textbf{89.36} $\pm$ 2.36 & \textbf{60.50} $\pm$ 6.18 \\ \hline
Zero-valued augmentation & 80.58 $\pm$ 3.86 & 78.45 $\pm$ 11.52 & 89.03 $\pm$ 2.38 & 59.09 $\pm$ 7.28 \\ \hline
Global covariance whitening & 79.46 $\pm$ 3.06 & 77.35 $\pm$ 11.59 & 88.56 $\pm$ 2.70 & 57.40 $\pm$ 5.78 \\ \hline
Classic MHA & 80.82 $\pm$ 3.40 & 76.96 $\pm$ 12.79 & 88.92 $\pm$ 2.09 & 60.16 $\pm$ 7.20 \\ \hline
Input length $L$ = 13 & 81.06 $\pm$ 3.49 & 78.79 $\pm$ 11.13 & 88.71 $\pm$ 2.31 & 60.39 $\pm$ 6.77 \\ \hline
Input length $L$ = 29 & 80.45 $\pm$ 3.87 & 77.70 $\pm$ 11.83 & 89.26 $\pm$ 2.36 & 59.57 $\pm$ 5.86 \\ \hline
\end{tabular}
\end{table}

\section{Comparison to the State-of-the-Art}
\label{sec:comparison}

To better contextualize our approach's capabilities, we test a number of previously published approaches (Section~\ref{ssec:SOA_scoring}) - more specifically, \textbf{DeepSleepNet} by Supratak et al.~\cite{supratak2017}, \textbf{IITNet} by Seo et al.~\cite{seo2020}, and the approach taken by Dequidt et al.~\cite{paul}, labeled \textbf{SleepVGG16} in this paper.

We compare them to the \textbf{SPDTransNet baseline}, as defined in Section~\ref{ssec:ablation_general}, as well as the variant of our model trained using learned augmentation features (Section~\ref{ssec:ablation_enrichment}), labeled \textbf{SPDTransNet+} in this section.

To ensure a fair comparison, we re-trained these models with our own methodology whenever possible, while keeping as much of the original authors' published implementation as possible.
Due to the absence of published hyperparameter research guidelines, we only use their published hyperparameters (or, if absent, those present in their published code) in our experiments.

Furthermore, DeepSleepNet already implements a rebalancing through oversampling in their pretrained epoch-wise component, before including it to an end-to-end sequence-to-sequence model. We have not modified this classification scheme.

Note that since we are optimizing for class-wise performance through the MF1 score when these approaches for the most part optimized the global (unweighted) accuracy score, our obtained results tend to be lower than those the original authors published.
This is true even for SleepVGG16, as even though the original approach also optimized through the MF1 score, it didn't include our test set clipping (Section~\ref{sec:method}).
We have found that this clipping tended to reduce the test set performance measures, presumably because epochs in the beginning and end of recordings (i.e. mostly corresponding to the Awake stage) are less ambiguous than other instances of the same stage.

\subsection{Learning From Scratch (LFS)}
\label{ssec:comparison_LFS}

For each dataset presented in Section~\ref{ssec:method_datasets}, we trained every considered model - i.e. both the baseline SPDTransNet model and the models presented in Section~\ref{sec:comparison}.
This allows us to directly compare our model's performance with State-of-the-Art approaches when using our class-wise performance optimization methodology.
The obtained results are displayed in Table~\ref{tab:comparison_LFS}.

As seen in the table, the size and relative homogeneity of MASS-SS3 (Section~\ref{ssec:method_datasets}) translated into the best performance overall for each considered model.
Interestingly, the higher degree of inter-subject variability in MASS-SS1 meant that even though this set has more than double the epoch count of Dreem DOD-H, the latter yielded better overall results than MASS-SS1, though the single-recording test sets led to a greater standard deviation in the aggregated results.

Model-wise, the baseline SPDTransNet performs similarly to or slightly better than SleepVGG16, with the learned augmentation variant outperforming both on SS3.
Both SPDTransNet and SleepVGG16 outperform DeepSleepNet and IITNet, though with a lead of around 3 points for SS3, 4 points for Dreem DOD-H and 6+ points for SS1.

This resiliency to increased inter-subject variability that both top models exhibit is most likely due to their common multi-signal approach, rather than the reliance on single-channel EEG analysis favored by DeepSleepNet and IITNet.

\begin{table}[]
\caption{Learning from scratch on considered datasets}
\label{tab:comparison_LFS}
\let\cline\clineorig
\setlength\tabcolsep{4pt}
\begin{tabular}{|c|c|c|c|c|c|}
\hline
Dataset & Model & MF1 & N3 F1 & N2 F1 & N1 F1 \\ \Xhline{3\arrayrulewidth}
\multirow{6}{*}{SS3} & DeepSleepNet & 78.14 $\pm$ 4.12 & 80.38 $\pm$ 9.35 & 89.25 $\pm$ 3.12 & 53.52 $\pm$ 8.24 \\ \cline{2-6} 
 & IITNet & 78.48 $\pm$ 3.15 & 81.97 $\pm$ 8.91 & 88.15 $\pm$ 2.84 & 56.02 $\pm$ 6.54 \\ \cline{2-6} 
 & SleepVGG16 & 81.23 $\pm$ 2.56 & \textbf{82.02} $\pm$ 8.76 & \textbf{90.57} $\pm$ 2.65 & 58.29 $\pm$ 5.01 \\ \cline{2-6} 
 & \textbf{SPDTransNet} & 81.24 $\pm$ 3.29 & 78.81 $\pm$ 11.17 & 89.36 $\pm$ 2.36 & \textbf{60.50} $\pm$ 6.18 \\ \cline{2-6} 
 & \textbf{SPDTransNet+} & \textbf{81.64} $\pm$ 2.88 & 80.15 $\pm$ 10.46 & 89.54 $\pm$ 2.46 & 60.29 $\pm$ 5.52 \\ \Xhline{3\arrayrulewidth}
\multirow{6}{*}{SS1} & DeepSleepNet & 68.49 $\pm$ 6.12 & 53.47 $\pm$ 21.59 & 83.74 $\pm$ 4.30 & 50.81 $\pm$ 8.19 \\ \cline{2-6} 
 & IITNet & 70.82 $\pm$ 6.09 & 56.40 $\pm$ 21.76 & 78.79 $\pm$ 6.20 & 52.29 $\pm$ 6.47 \\ \cline{2-6} 
 & SleepVGG16 & 77.31 $\pm$ 5.46 & \textbf{66.75} $\pm$ 20.76 & \textbf{86.51} $\pm$ 3.82 & 62.59 $\pm$ 5.34 \\ \cline{2-6} 
 & \textbf{SPDTransNet} & \textbf{77.75} $\pm$ 5.41 & 63.08 $\pm$ 20.96 & 85.28 $\pm$ 4.55 & \textbf{63.29} $\pm$ 6.02 \\ \Xhline{3\arrayrulewidth}
\multirow{6}{*}{DOD-H} & DeepSleepNet & 73.07 $\pm$ 13.66 & 75.98 $\pm$ 23.58 & 84.50 $\pm$ 15.20 & 48.40 $\pm$ 16.86 \\ \cline{2-6} 
 & IITNet & 73.55 $\pm$ 9.36 & 76.67 $\pm$ 24.21 & 86.32 $\pm$ 6.27 & 49.68 $\pm$ 12.53 \\ \cline{2-6} 
 & SleepVGG16 & 75.80 $\pm$ 10.60 & 76.87 $\pm$ 24.15 & 85.99 $\pm$ 11.72 & 51.30 $\pm$ 12.00 \\ \cline{2-6} 
 & \textbf{SPDTransNet} & \textbf{77.48} $\pm$ 8.74 & \textbf{77.75} $\pm$ 24.04 & \textbf{87.40} $\pm$ 5.42 & \textbf{56.99} $\pm$ 11.93 \\ \hline
\end{tabular}
\end{table}

\subsection{Direct Transfer (DT)}
\label{ssec:comparison_DT}

To be used in a clinical setting, a sleep stage scoring model would need to retain a good level of performance when utilized with newly acquired data.
As such, we have elected to compare the ability of all considered models to adapt to other datasets.
These experiments are not undertaken using the Dreem DOD-H dataset, as it differs from both SS1 and SS3 in the number and nature of the considered electrodes.

For each model, we utilize the LFS weights obtained on each fold of MASS-SS3 (Section~\ref{tab:comparison_LFS}), and test on all recordings of the MASS-SS1 dataset.
The results are displayed in Table~\ref{tab:comparison_DT}.

One first observation is that for each model, the standard deviations tend to be lower than their LFS equivalent, both on SS1 and SS3, particularly for our model and SleepVGG16.
As stated in Section~\ref{sec:ablation}, the high standard deviations in Table~\ref{tab:comparison_LFS} are a consequence of our cross-validation strategy.
Since test sets between folds are relatively small and non-overlapping, inter-test-set variability tends to be high, inflating the resulting standard deviation when aggregating results.
By using a unified test set for each fold, we remove this factor entirely, yielding lower standard deviations than for the LFS experiments, and increasing our confidence in the significance of the obtained results.

The most striking result, however, is the dramatic reduction in performance between LFS and DT results for both single-channel models, with a loss of 15+ points on DeepSleepNet and IITNet compared to their LFS performance on MASS-SS1 - not to mention their relatively high standard deviations, indicative of inter-fold instability.

By contrast, SleepVGG16 lost slightly less than 3 points, and the baseline SPDTransNet, less than one point.
Hence, it would seem that our approach through functional connectivity does outperform SleepVGG16 in resiliency when in a multi-dataset context.

Interestingly, the SPDTransNet+ configuration outperforms even our baseline LFS results on SS1.
This would indicate that the feature extraction through 1D CNN strategy utilized by DeepSleepNet and IITNet isn't the main source of their lack of flexibility, as SPDTransNet+ was able to utilize the epoch-wise feature extraction submodel of IITNet (Section~\ref{ssssec:model_data_enrichment_augmentation}) to great extent.
Rather, it would seem that said inflexibility is once again due to their monosignal nature.

\begin{table}[]
\centering
\caption{Direct transfer, i.e. testing models trained on MASS-SS3 with recordings from another dataset}
\label{tab:comparison_DT}
\begin{tabular}{|c|c|c|c|c|}
\hline
Model & MF1 & N3 F1 & N2 F1 & N1 F1 \\ \hline
DeepSleepNet & 52.41 $\pm$ 3.15 & 55.43 $\pm$ 5.95 & 61.17 $\pm$ 6.44 & 32.11 $\pm$ 3.75 \\ \hline
IITNet & 55.27 $\pm$ 4.08 & 61.05 $\pm$ 4.39 & 69.82 $\pm$ 6.71 & 36.54 $\pm$ 4.42 \\ \hline
SleepVGG16 & 74.68 $\pm$ 1.82 & 70.82 $\pm$ 3.80 & \textbf{86.72} $\pm$ 0.62 & 55.79 $\pm$ 2.36 \\ \hline
\textbf{SPDTransNet} & 76.99 $\pm$ 0.94 & 67.59 $\pm$ 2.08 & 86.52 $\pm$ 0.54 & 57.92 $\pm$ 1.85 \\ \hline
\textbf{SPDTransNet+} & \textbf{78.23} $\pm$ 0.62 & \textbf{68.71} $\pm$ 2.19 & 86.43 $\pm$ 0.91 & \textbf{59.68} $\pm$ 1.72 \\ \hline
\end{tabular}
\end{table}

\subsection{Finetuning (FT)}
\label{ssec:comparison_FT}

In addition to DT (Section~\ref{ssec:comparison_DT}), and for the same reasons, we tested the considered models' ability to improve classification through transfer learning, loading weights trained on MASS-SS3 (Section~\ref{ssec:comparison_LFS}) and finetuning them by re-training on each fold of the other considered datasets.
For the MASS-SS3 weights, we selected those trained using the fold used in hyperparameter researches (Section~\ref{ssec:method_datasets}).

As seen in Table~\ref{tab:comparison_FT}, in this configuration, SPDTransNet is able to improve its classification performance when compared to learning from scratch on MASS-SS1, with SleepVGG16 not far behind, and the SPDTransNet+ variant once again outperforming both.
However, the same cannot be said for DeepSleepNet and IITNet, whose performance was comparable to the one obtained in the LFS configuration - meaning that prior learning on SS3 had the same impact as random weight initialization when it comes to learn on SS1.

Combined with the results from Section~\ref{ssec:comparison_DT}, this tells us that differences between EEG sleep staging datasets can be large enough to make non-LFS learning strategies irrelevant, particularly for models trained on few EEG signals.
However, multi-signal models, such as our SPDTransNet network and SleepVGG16, have proven to be robust to such changes, allowing it them retain decent performance in DT and  even improve in FT.
Additionally, SPDTransNet's learned augmentation variant has proved particularly well-performing across the board, with the handcrafted augmentation baseline performing on par with the SleepVGG16 model, and outperforming it when exposed to an unknown dataset.

\begin{table}[]
\centering
\caption{Finetuning weights trained on MASS-SS3 using another dataset}
\label{tab:comparison_FT}
\begin{tabular}{|c|c|c|c|c|}
\hline
Model & MF1 & N3 F1 & N2 F1 & N1 F1 \\ \hline
DeepSleepNet & 67.65 $\pm$ 7.47 & 55.15 $\pm$ 22.58 & 81.03 $\pm$ 5.42 & 47.48 $\pm$ 7.03 \\ \hline
IITNet & 70.95 $\pm$ 5.85 & 56.51 $\pm$ 21.63 & 80.46 $\pm$ 6.46 & 53.85 $\pm$ 5.90 \\ \hline
SleepVGG16 & 78.90 $\pm$ 4.61 & \textbf{68.88} $\pm$ 18.53 & \textbf{86.41} $\pm$ 3.91 & 64.15 $\pm$ 6.19 \\ \hline
\textbf{SPDTransNet} & 78.99 $\pm$ 4.45 & 64.65 $\pm$ 18.69 & 86.19 $\pm$ 5.09 & 64.60 $\pm$ 5.46 \\ \hline
\textbf{SPDTransNet+} & \textbf{79.41} $\pm$ 4.40 & 65.90 $\pm$ 18.21 & 85.90 $\pm$ 5.02 & \textbf{65.16} $\pm$ 6.24 \\ \hline
\end{tabular}
\end{table}

\section{Conclusion}
\label{sec:conclusion}

In this paper, we have explored the capabilities of our SPDTransNet model, a Transformer-based model designed for the structure-preserving analysis of timeseries of Symmetric Positive Definite covariance matrices, in the classification of sleep stages from a collection of electroencephalographic signals.

Beyond once again proving the relevance of an analysis through functional connectivity for this task, we have showcased our model's flexibility in multi-dataset environment.
In particular, we outperformed an otherwise equivalent model when classifying signals from an unknown dataset, with minimal loss of performance compared to training on said dataset.

Furthermore, we have expanded upon our signal processing pipeline, enriching our covariance matrices through the inclusion of learned signal-specific features, further improving both our model's baseline performance and flexibility.

\backmatter

\bmhead{Supplementary Material}

Specifics regarding some portions of this paper have been included as Supplementary Material, and are located at the end of this document.

\section*{Statements and Declarations}

\bmhead{Competing Interests}

The authors declare that they have no conflict of interest.

\bmhead{Funding Information}

This work was granted access to computing resources of IDRIS and CRIANN (see Acknoledgements).
The first author is supported by the French National Research Agency (ANR) and Region Normandie under grant HAISCoDe.







\bmhead{Author contribution}

All authors contributed to the study conception and design. Experiments were performed by Mathieu Seraphim, under the supervision of Alexis Lechervy, Luc Brun and Olivier Etard. The first draft of the manuscript was written by Mathieu Seraphim and all authors commented on previous versions of the manuscript. All authors read and approved the final manuscript.

\bmhead{Data Availability}

Our code is publicly available in the following repository: \url{github.com/MathieuSeraphim/SPDTransNet_plus}.
The Montreal Archive of Sleep Studies (MASS) dataset is available upon request, pending approval from the Ethics Review Board (ERB) of the MASS host~\cite{MASS}.
The Dreem DOD-H dataset is freely available online, and downloading instructions can be found in the authors' GitHub repository~\cite{guillot2020}.

\bmhead{Research Involving Human and /or Animals}

Both utilized datasets are composed of anonymized recordings of human biological signals.
The MASS recordings were anonymized and distributed in accordance with guidelines from the MASS research team's ethics board~\cite{MASS}.
The Dreem DOD-H dataset was compiled and made publicly available with approval from the Committees of Protection of Persons (CPP), and declared and carried out in accordance with French law~\cite{guillot2020}.
No further biological data was utilized in the production of this paper.

\bmhead{Informed Consent}

Not applicable (see above).

\bibliography{biblio}

\newpage

\part*{Supplementary Material}

\setcounter{section}{0}
\renewcommand*{\theHsection}{chX.\the\value{section}}
\setcounter{figure}{0}
\renewcommand\thefigure{\arabic{figure}}
\setcounter{table}{0}
\renewcommand\thetable{\arabic{table}}   

\section{Proof of SPD Preservation through Augmentation}
\label{supsec:augmentation}

Let $X$ $\in$ $SPD(n)$, $A$ $\in$ $\mathbb{R}^{n \times k}$ and $x$ $\in$ $\mathbb{R}^{n+k}$ be arbitrary elements of their respective set.
In addition, let $x_1$ $\in$ $\mathbb{R}^n$ and $x_2$ $\in$ $\mathbb{R}^k$ be defined so that:
\[x^T = (x_1^T,\ x_2^T)\]
In other words, $x$ is the concatenation of $x_1$ and $x_2$.

Let $X'$ = $aug_\alpha(X, A)$ $\in$ $\mathbb{R}^{m \times m}$, with $m$ = $n+k$, be the result of the augmentation of $X$ by $A$ (Section~\ref{ssec:SPD_augmentation} of the paper).
Since we make no assumption on the value taken by $A$, we set $\alpha$ = 1, without loss of generality\footnote{If $\alpha$ $\not\in$ \{0, 1\}, replace $A$ with $A'$ = $\frac{1}{\alpha} \cdot A$. If $\alpha$ = 0, replace it with $0_{n \times k}$.}.\\

By definition,
\[X' \in SPD(m) \Leftrightarrow \forall x \neq 0_{m},\ x^T \cdot X' \cdot x > 0\]
with $0_{m}$ $\in$ $\mathbb{R}^{m}$ a vector of zeros.\\

\[x^T \cdot X' \cdot x = (x_1^T,\ x_2^T) \cdot 
\left(\begin{array}{c|c}
\\
X + A \cdot A^T & A \\
\\
\hline
A^T & \mathbb{I}_k \\
\end{array}\right) \cdot
\left(\begin{array}{c}
x_1\\
x_2\\
\end{array}\right)\]
\[= (x_1^T \cdot X + x_1^T \cdot A \cdot A^T + x_2^T \cdot A^T,\ x_1^T \cdot A + x_2^T \cdot \mathbb{I}_k) \cdot
\left(\begin{array}{c}
x_1\\
x_2\\
\end{array}\right)\]
\[= x_1^T \cdot X \cdot x_1 + x_1^T \cdot A \cdot A^T \cdot x_1 + x_2^T \cdot A^T \cdot x_1 + x_1^T \cdot A \cdot x_2 + x_2^T \cdot \mathbb{I}_k \cdot x_2\]
\[= x_1^T \cdot X \cdot x_1 + (x_1^T \cdot A) \cdot (x_1^T \cdot A)^T + (x_1^T \cdot A) \cdot x_2 + x_2^T \cdot (x_1^T \cdot A)^T + x_2^T \cdot x_2\]

Therefore:
\[x^T \cdot X' \cdot x = x_1^T \cdot X \cdot x_1 + (x_1^T \cdot A + x_2^T) \cdot (x_1^T \cdot A + x_2^T)^T\]
with $x_1^T \cdot A$ and $x_2^T$ line vectors of length $k$.
\begin{itemize}
    \item[] $\rightarrow (x_1^T \cdot A + x_2^T) \cdot (x_1^T \cdot A + x_2^T)^T \geq 0$
\end{itemize}

However:
\begin{itemize}
    \item If $x_1 \neq 0_n$, $x_1^T \cdot X \cdot x_1 > 0$ (by definition, as $X$ $\in$ $SPD(n)$)
    \item If $x_1 = 0_n$ but $x_2 \neq 0_k$, $(x_1^T \cdot A + x_2^T) \cdot (x_1^T \cdot A + x_2^T)^T = x_2^T \cdot x_2 > 0$
\end{itemize}~\\

In conclusion, whichever the values of $x_1$ and $x_2$, and therefore for any arbitrary vector $x$ $\in$ $\mathbb{R}^{m}$, we have:
\[x^T \cdot X' \cdot x = 0 \Leftrightarrow x = 0_{m}\]
Which proves that $X'$ = $aug_1(X, A)$ is SPD, and more generally, that the augmentation operation presented in Equation~\ref{eq:augmentation} of the paper preserves the SPD nature of its input $X$.

\section{The Signal-Wise Feature Extraction Submodel}
\label{supsec:IITNet}

As stated in Section~\ref{ssssec:model_data_enrichment_augmentation} of the paper, we utilize the epoch-wise feature extraction CNN utilized by Seo et al. in IITNet\cite{seo2020}.
It is derived from the ResNet-50 architecture, modified to analyze 1D signals, and which outputs $N$ feature vectors of length 128.
Each feature vector is derived from a section of the input signal, with the number of vectors $N$ and their receptive fields being determined by the length of the input signal. Since EEG epochs have a fixed duration of 30 seconds (Section~\ref{ssec:SOA_scoring} of the paper), these will depend on the sampling frequency of the signals.

As seen in Seo et al.'s original publication, an input of length 3000 (corresponding to an epoch sampled at 100Hz) results in $N$ = 47 feature vectors of $\mathbb{R}^{128}$. With MASS-SS1 and MASS-SS3, both sampled at 256Hz, this becomes $N$ = 120.
Given our subdivision of each epoch in 30 one-second segments (Section~\ref{sssec:model_data_covariances} of the paper), this equates to 4 vectors of $\mathbb{R}^{128}$ (i.e. 512 features) per segment, channel and signal.

\begin{figure}[h]
\centering
\includegraphics[width=1\textwidth]{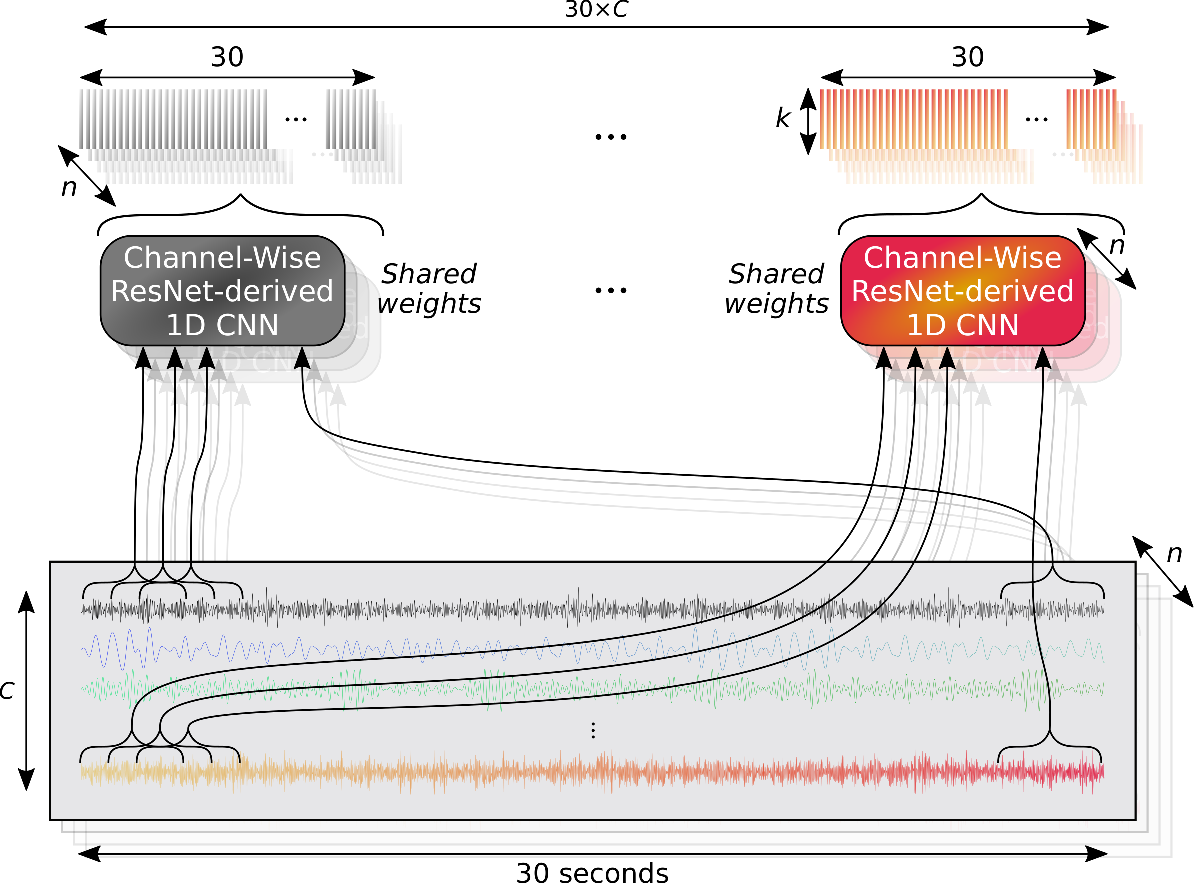}
\caption{Epoch-wise feature extraction submodel of IITNet~\cite{seo2020}, adapted to extract augmentation features - with $n$ the number of EEG signals, $C$ the number of channels, and $k$ the size of each signal-wise feature vector.}
\label{fig:learned_augmentation}
\end{figure}

We modify this CNN in two ways.
Firstly, we adapt it to handle multi-channel, multi-signal inputs.
As seen in Figure~\ref{fig:learned_augmentation}, for each epoch, features from each EEG signal and each channel are computed in parallel, sharing weights between signals in the same channel, but not between channels.
Secondly, we reduce the 512 features per one-second segment into $k$ features through a fully connected layer, obtaining a timeseries of 30 vectors of $\mathbb{R}^k$ per channel and signal.

Combining the $n$ signals for a given segment and channel, we obtain the $n \times k$ feature matrix that we use to augment our covariance matrices.
As seen in Table~\ref{tab:receptive_field}, the receptive field corresponding to each feature vector of $\mathbb{R}^k$ always includes the time segment on which the corresponding matrix is computed, with additional overlap with neighboring segments.

\begin{table}[h]
\caption{Receptive fields for each of the 30 augmentation feature vectors extracted, for epochs sampled at 256 Hz.}
\label{tab:receptive_field}
\begin{tabular}{|c|c|c|c|c|}
\hline
ID & Lower sample & Upper sample & Starting time (s) & Duration (s) \\ \hline
0 & 0 & 589 & 0 & 2.3 \\ \hline
1 & 0 & 845 & 0 & 3.3 \\ \hline
2 & 131 & 1101 & 0.51 & 3.79 \\ \hline
3 & 387 & 1357 & 1.51 & 3.79 \\ \hline
4 & 643 & 1613 & 2.51 & 3.79 \\ \hline
... & ... & ... & ... & ... \\ \hline
25 & 6019 & 6986 & 23.51 & 3.79 \\ \hline
26 & 6275 & 7245 & 24.51 & 3.79 \\ \hline
27 & 6531 & 7501 & 25.51 & 3.79 \\ \hline
28 & 6787 & 7679 & 26.51 & 3.48 \\ \hline
29 & 7043 & 7679 & 27.51 & 2.48 \\ \hline
\end{tabular}
\end{table}

\end{document}